\documentclass{article}
\usepackage{amssymb,amsmath}
\newtheorem{theorem}{Theorem}[section]
\newtheorem{definition}{Definition}[section]
\newtheorem{remark}{Remark}[section]
\newtheorem{conjecture}{Conjecture}[section]
\newtheorem{lemma}{Lemma}[section]

\begin{document}

\title{Pseudoinstantons in metric-affine field theory}
\author{Dmitri Vassiliev
\thanks{Department of Mathematical Sciences, University of Bath,
Bath BA2 7AY, UK.
Email D.Vassiliev@bath.ac.uk,
URL http://www.bath.ac.uk/\~{}masdv/}
}
\maketitle

\begin{abstract}
In abstract Yang--Mills theory the standard instanton construction
relies on the Hodge star having real eigenvalues which makes it
inapplicable in the Lorentzian case.
We show that for the affine connection
an instanton-type construction can be carried out
in the Lorentzian setting.
The Lorentzian analogue of an instanton is a spacetime
whose connection is metric compatible and Riemann curvature
irreducible (``pseudoinstanton'').
We suggest a metric-affine action which is a natural generalization
of the Yang--Mills action and for which
pseudo\-instantons are stationary points.
We show that a spacetime with a Ricci flat Levi-Civita connection
is a pseudoinstanton, so
the vacuum Einstein equation is a special case of our theory.
We also find another pseudoinstanton
which is a wave of torsion in Minkowski space. Analysis
of the latter solution indicates the possibility of using it
as a model for the neutrino.
\end{abstract}


\newpage
\tableofcontents
\newpage

\section{Statement of the problem}
\label{statement}

We consider spacetime to be a connected real oriented
4-manifold $M$ equipped with a Lorentzian metric $g$
and an affine connection $\Gamma$.
The 10 independent
components of the metric tensor $g_{\mu\nu}$
and the 64 connection coefficients ${\Gamma^\lambda}_{\mu\nu}$
are the unknowns of our theory, as is the manifold $M$ itself.

It is known (see Appendix B.4 in \cite{hehlreview}
as well as Appendix \ref{appendixa} in our paper)
that at each point $x\in M$ the vector space
of (real) Riemann curvatures decomposes under the Lorentz group into
a direct sum of eleven invariant subspaces
which are irreducible and mutually orthogonal.
Given a Riemann curvature $R$ we will denote by $R^{(j)}$,
$j=1,\ldots,11$, its irreducible pieces.

The natural inner product on Riemann curvatures is
\[
(R,Q):=\int R^\kappa{}_{\lambda\mu\nu}\,
\,Q^\lambda{}_\kappa{}^{\mu\nu}\,.
\]
We denote $\|R\|^2:=(R,R)$.
Of course, our inner product is indefinite,
so $\|R\|^2$ does not have a particular sign and
we cannot attribute a meaning to $\|R\|$ itself.

We define our action as
\begin{equation}
\label{action}
S:=\sum_{j=1}^{11}c_j\,\|R^{(j)}\|^2,
\end{equation}
where the $c_j$'s are real constants.
Note the analogy between formula (\ref{action}) and
the potential energy of an isotropic elastic body,
see formulae (4.2), (4.3) in \cite{LL7}.
The only difference is that in the theory of elasticity the
field strength is the deformation tensor
(rank 2 symmetric tensor)
rather than Riemann curvature, and it has two
irreducible pieces (shear and hydrostatic compression) rather than eleven.
Note also that the idea of using an action of the type
(\ref{action}) goes back to
Weyl who argued
at the end of his 1919 paper
\cite{weylquadraticaction}
that the most natural
gravitational action should be quadratic in curvature and
involve its irreducible pieces as separate terms.
Weyl wrote:
``I intend to pursue the consequences of this action principle
in a continuation of this paper''.
(Translation by G.~Friesecke.)
It is regrettable that Weyl never
carried out this analysis.

Variation of the action (\ref{action}) with respect to the metric
$g$ and the connection $\Gamma$ produces Euler--Lagrange
equations which we will write symbolically as
\begin{eqnarray}
\label{eulerlagrangemetric}
\partial S/\partial g&=&0,
\\
\label{eulerlagrangeconnection}
\partial S/\partial\Gamma&=&0.
\end{eqnarray}
Our objective is the study of the
combined system
(\ref{eulerlagrangemetric}), (\ref{eulerlagrangeconnection}).
This is a system of $10+64$
real nonlinear partial differential equations
with $10+64$ real unknowns.

\begin{remark}
\label{conformal}
It is easy to see that
the action (\ref{action}) is conformally invariant,
i.e, it does not change if we perform a Weyl rescaling of the metric
$g\to e^{2f}g$, $f:M\to\mathbb{R}$,
without changing the connection $\Gamma$
(here it is important that in the metric-affine setting
the metric and the connection lead a separate existence).
Therefore, the number of independent equations in
(\ref{eulerlagrangemetric}) is not 10 but 9.
\end{remark}

Following Eisenhart \cite{eisenhart}
we call a spacetime \emph{Riemannian} if its connection is Levi-Civita
(i.e., ${\Gamma^\lambda}_{\mu\nu}=
\genfrac{\{}{\}}{0pt}{}{\lambda}{\mu\nu}$)
and  \emph{non-Riemannian} otherwise.
Here ``Riemannian'' does not imply
the positivity of the metric,
the latter being assumed to be Lorentzian throughout the paper.

In the special case
\begin{equation}
\label{specialcase}
c_1=\ldots=c_{11}=1
\end{equation}
the functional (\ref{action})
becomes $\|R\|^2$.
This is the Yang--Mills action for the affine
connection, and
equation (\ref{eulerlagrangeconnection}) is the corresponding
Yang--Mills equation. The latter was analyzed by Yang \cite{yang}.
Yang was looking for Riemannian solutions,
so he specialized equation
(\ref{eulerlagrangeconnection}) to the Levi-Civita
connection and arrived at the equation
\begin{equation}
\label{compact}
\nabla_\lambda Ric_{\kappa\mu}
-\nabla_\kappa Ric_{\lambda\mu}=0.
\end{equation}
Here ``specialization'' means that one sets
${\Gamma^\lambda}_{\mu\nu}=\genfrac{\{}{\}}{0pt}{}{\lambda}{\mu\nu}$
\emph{after} the variation in $\Gamma$ is carried out.
An immediate consequence of equation (\ref{compact})
is the fact that Einstein spaces
satisfy the Yang--Mills equation (\ref{eulerlagrangeconnection}).

A number of other authors observed, still under the assumption
(\ref{specialcase}), that a much stronger result is true: Einstein
spaces satisfy both equations (\ref{eulerlagrangemetric}) and
(\ref{eulerlagrangeconnection}). An elegant explanation of this
fact in terms of double duality was given by Mielke
\cite{mielkepseudoparticle}. Mielke's paper was written for the
case of a positive metric but the result remains true for the
Lorentzian case, the only difference being that one has to change
signs in double duality formulae. We shall therefore refer to the
special case (\ref{specialcase}) of the model
(\ref{eulerlagrangemetric}), (\ref{eulerlagrangeconnection})
as the \emph{Yang--Mielke theory of gravity}.

Apart from \cite{yang,mielkepseudoparticle} there have
been numerous other publications on the subject,
with many authors independently rediscovering known results.
One can get an idea of
the historical development
of the Yang--Mielke theory of gravity from
\cite{stephenson,
buchdahl,
thompsonFeb1975,
pavelleApr1975,
thompsonAug1975,
fairchild1976,
fairchild1976erratum,
olesen,
wilczek}.
Of these publications the most remarkable is
the Mathematical Review \cite{buchdahl}:
the author of the review
noticed a fact missed in the
paper under review \cite{stephenson}, namely,
that Einstein spaces are stationary points of
the Yang--Mills action with respect to the variation of
both the metric and the connection,
a fact repeatedly rediscovered in later years.

Our aim is to develop the Yang--Mielke theory of gravity by
\begin{itemize}
\item
dropping the requirement (\ref{specialcase}),
\item
looking for Riemannian solutions other than
Einstein spaces, and
\item
looking for non-Riemannian solutions.
\end{itemize}

\section{Main result}
\label{mainresult}

The following definition is crucial in our construction.

\begin{definition}
\label{definitionofapseudoinstanton}
We call a spacetime a \emph{pseudoinstanton}
if its connection is metric compatible and Riemann curvature
irreducible.
\end{definition}

Here irreducibility of Riemann curvature
means that of the eleven $R^{(j)}$'s all except one
are identically zero.
In fact, metric compatibility cuts the number of possible irreducible
pieces to six. Explicit formulae for the latter
are given at the end of Appendix \ref{appendixa}.

Definition \ref{definitionofapseudoinstanton} is motivated
by the analogy with abstract Yang--Mills theory in Euclidean space,
see Sections 3 and 4 of Chapter 1 in \cite{atiyah}.
Indeed, the notion of an instanton is based
on the decomposition of the vector space of curvatures
into two subspaces which are invariant under
the action of the orthogonal group on the external indices.
(We call the Lie algebra indices of curvature internal,
and the remaining ones external.) The case of the
affine connection is special in that the internal and
external indices have the same nature, so
it is logical to apply (pseudo)orthogonal transformations
to the whole rank 4 tensor.
This leads to a richer algebraic structure.

Our main result is

\begin{theorem}
\label{maintheorem}
A pseudoinstanton is a solution of
the problem
(\ref{eulerlagrangemetric}), (\ref{eulerlagrangeconnection}).
\end{theorem}

In Section \ref{proofofmaintheorem} we prove Theorem \ref{maintheorem},
and in the remainder of the paper we use this theorem
for constructing families of solutions of the system
(\ref{eulerlagrangemetric}), (\ref{eulerlagrangeconnection}).

\section{Notation}
\label{notation}

Our notation follows \cite{kingandvassiliev}. In particular,
we denote local coordinates by $x^\mu$, $\mu=0,1,2,3$,
and write $\partial_\mu:=\partial/\partial x^\mu$.
We define the covariant derivative of a vector function as
$\nabla_{\mu}v^\lambda:=\partial_\mu v^\lambda
+{\Gamma^\lambda}_{\mu\nu}v^\nu$,
torsion as
${T^\lambda}_{\mu\nu}:=
{\Gamma^\lambda}_{\mu\nu}-{\Gamma^\lambda}_{\nu\mu}\,$,
contortion as
\begin{equation}
\label{contortionviatorsion}
{K^\lambda}_{\mu\nu}:=\frac12
\bigl(
T{}^\lambda{}_{\mu\nu}+T{}_\mu{}^\lambda{}_\nu+T{}_\nu{}^\lambda{}_\mu
\bigr)
\end{equation}
(see formula (7.35) in \cite{nakahara}),
Riemann curvature as
\begin{equation}
\label{riemanncurvature}
{R^\kappa}_{\lambda\mu\nu}:=
\partial_\mu{\Gamma^\kappa}_{\nu\lambda}
-\partial_\nu{\Gamma^\kappa}_{\mu\lambda}
+{\Gamma^\kappa}_{\mu\eta}{\Gamma^\eta}_{\nu\lambda}
-{\Gamma^\kappa}_{\nu\eta}{\Gamma^\eta}_{\mu\lambda}\,,
\end{equation}
Ricci curvature as
$Ric_{\lambda\nu}:={R^\kappa}_{\lambda\kappa\nu}\,$,
scalar curvature as $\mathcal{R}:=Ric^\lambda{}_\lambda\,$,
and trace free Ricci curvature as
$\mathcal{R}ic_{\lambda\nu}:=
Ric_{\lambda\nu}-\frac14g_{\lambda\nu}\mathcal{R}\,$.
We denote Weyl curvature by $\mathcal{W}=R^{(3)}$.

It is easy to see that contortion
has the antisymmetry property
$K_{\lambda\mu\nu}=-K_{\nu\mu\lambda}$
and that
\begin{equation}
\label{torsionviacontortion}
{T^\lambda}_{\mu\nu}=
{K^\lambda}_{\mu\nu}-{K^\lambda}_{\nu\mu}\,.
\end{equation}
Formulae
(\ref{contortionviatorsion}), (\ref{torsionviacontortion})
allow us to express torsion and contortion via
one another.

A connection is said to be metric compatible if
$\nabla_\lambda g_{\mu\nu}\equiv0$.
A metric compatible connection is uniquely determined
by metric and torsion or metric and contortion,
see Section 7.2.6 in \cite{nakahara} for details.
In the metric compatible case contortion can be written as
\begin{equation}
\label{contortionviagammaandchristoffel}
{K^\lambda}_{\mu\nu}=
{\Gamma^\lambda}_{\mu\nu}-\genfrac{\{}{\}}{0pt}{}{\lambda}{\mu\nu},
\end{equation}
where
\begin{equation}
\label{christoffel}
\genfrac{\{}{\}}{0pt}{}{\lambda}{\mu\nu}:=
\frac12g^{\lambda\kappa}
(\partial_\mu g_{\nu\kappa}
+\partial_\nu g_{\mu\kappa}
-\partial_\kappa g_{\mu\nu})
\end{equation}
is the Christoffel symbol.

The choice between using torsion and
using contortion is a matter of taste.
When working with metric compatible connections using contortion
is somewhat more convenient because formula
(\ref{contortionviagammaandchristoffel})
is so simple and natural.

Given a scalar function $f:M\to\mathbb{R}$ we write for brevity
\[
\int f:=\int_Mf\,\sqrt{|\det g|}
\ dx^0dx^1dx^2dx^3,
\qquad
\det g:=\det(g_{\mu\nu}).
\]

Throughout the paper we work only in coordinate systems with
positive orientation. Moreover, when we restrict our consideration
to Minkowski space we assume that our coordinate frame is obtained
from a given reference frame by a proper Lorentz transformation.

We define the action of the Hodge star on
a rank $q$ antisymmetric tensor as
\begin{equation}
\label{hodgestar}
(*Q)_{\mu_{q+1}\ldots\mu_4}\!:=(q!)^{-1}\,\sqrt{|\det g|}
\ Q^{\mu_1\ldots\mu_q}\varepsilon_{\mu_1\ldots\mu_4}\,,
\end{equation}
where $\varepsilon$ is the totally antisymmetric quantity,
$\varepsilon_{0123}:=+1$.

\section{Proof of the main theorem}
\label{proofofmaintheorem}

This section is devoted to the proof of
Theorem \ref{maintheorem}.

Let us first examine what happens when we fix the metric
and vary the connection. The explicit formula for the variation
of the action $\delta S$ resulting from the variation of the
connection $\delta\Gamma$ is
\begin{multline*}
\delta S
=2\sum_{j=1}^{11}c_j\int{\rm tr}\bigl(
(R^{(j)})^{\mu\nu}(\delta R^{(j)})_{\mu\nu}
\bigr)
=2\sum_{j=1}^{11}c_j\int{\rm tr}\bigl(
(R^{(j)})^{\mu\nu}(\delta R)_{\mu\nu}
\bigr)
\\
=4\sum_{j=1}^{11}c_j\int{\rm tr}\bigl(
(\delta_{{\rm YM}}R^{(j)})^\mu(\delta\Gamma)_\mu
\bigr)
\end{multline*}
where $\delta_{{\rm YM}}$ is the Yang--Mills divergence,
\[
(\delta_{\rm YM}R)^\mu:=\frac1{\sqrt{|{\det}g|}}\,
(\partial_\nu+[\Gamma_\nu,\,\cdot\,])
(\sqrt{|{\rm det}g|}\,R^{\mu\nu})\,.
\]
Here, as in \cite{kingandvassiliev,kiev}, we use matrix notation
to hide the two internal indices.
We start our variation from a spacetime with a metric
compatible connection
(see Definition \ref{definitionofapseudoinstanton})
and this fact has important consequences.
We have $R^{(j)}\equiv0$ for $j=7,\ldots,11$
(see Appendix \ref{appendixa} for details).
The remaining curvatures $R^{(j)}$, $j=1,\ldots,6$,
are antisymmetric in the internal indices and, moreover,
the action of the Yang--Mills divergence preserves this property.
(This is, of course,
a consequence of the fact that antisymmetric rank 2
tensors form a subalgebra within the general
Lie algebra of rank 2 tensors.)
Therefore, in order to prove that we have a stationary
point with respect to arbitrary variations of the connection
it is necessary and sufficient to prove that
we have a stationary
point with respect to variations of the connection
which are antisymmetric in the internal indices, i.e.,
variations satisfying
$g_{\kappa\lambda}(\delta\Gamma)^\lambda{}_{\mu\nu}
+
g_{\nu\lambda}(\delta\Gamma)^\lambda{}_{\mu\kappa}=0$.
But this means that
it is necessary and sufficient to prove that
we have a stationary
point with respect to variations of the connection which
preserve metric compatibility.
So further on in this section
we work with metric compatible connections only.

We start our variation from a spacetime which is a pseudoinstanton,
therefore for some $l\in\{1,\ldots,6\}$ we have
\begin{equation}
\label{detailedproof1}
R^{(j)}\equiv0,\quad\forall j\ne l.
\end{equation}
Let us rewrite formula (\ref{action}) as
\begin{equation}
\label{detailedproof2}
S=c_l\|R\|^2+
\sum_{j=1}^{11}(c_j-c_l)\|R^{(j)}\|^2
\end{equation}
and vary the metric and the connection.
Formulae (\ref{detailedproof1}), (\ref{detailedproof2})
imply that $\delta S=c_l\,\delta(\|R\|^2)$.
So in order to prove Theorem \ref{maintheorem}
it is sufficient to show that our pseudoinstanton is
a stationary point of the Yang--Mills action $\|R\|^2$.

The remainder of the proof is an adaptation
of Mielke's argument \cite{mielkepseudoparticle}.

Let us first assume for simplicity that our manifold
is compact.
For compact manifolds with metric compatible
connections we have the identity
\begin{equation}
\label{proofofmaintheorem1}
\|R\|^2=\frac12\|R\mp{}^*\!R^*\|^2\pm(R,{}^*\!R^*),
\end{equation}
where ${}^*\!R^*$ is defined in accordance
with formulae
(\ref{appendixa2}), (\ref{appendixa3}), (\ref{doubleduality}).
It is known that $(R,{}^*\!R^*)$ is a topological invariant:
it is, up to a normalizing factor, the Euler number of the manifold,
see Section 5 of Chapter XII and Note 20 in \cite{KN2}.
(Actually, the Euler number of
a compact Lorentzian manifold can only be
zero, see \cite{steenrod}, p. 207.)
So it
remains to show that a pseudoinstanton is a stationary point
of the functional $\|R\mp{}^*\!R^*\|^2$. But this is a consequence
of the fact that
irreducibility of Riemann curvature implies ${}^*\!R^*=\pm R$,
see Appendix \ref{appendixa} for details.

In the case of a noncompact manifold one should understand
the identity
(\ref{proofofmaintheorem1}) in the
Euler--Lagrange sense. The statement that
$(R,{}^*\!R^*)$ is a topological invariant
means now that this functional generates zero Euler--Lagrange
terms. Euler--Lagrange arguments are purely local
and the fact that $(R,{}^*\!R^*)$ does not contribute to the
Euler--Lagrange equations is unrelated to the compactness
or noncompactness of the manifold.

The proof of Theorem \ref{maintheorem} is complete.

\section{The Weyl pseudoinstanton}
\label{weylpseudoinstanton}

In order to start constructing pseudoinstantons we need
to choose the irreducible subspace $\mathbf{R}^{(l)}$
into which we will attempt to fit our Riemann curvature.
We choose to look first for pseudoinstantons
in the subspace of Weyl curvatures
$\mathbf{R}^{(3)}$.
This choice is motivated by the observation that of the
six possible subspaces
$\mathbf{R}^{(l)}$, $l=1,\ldots,6$,
generated by a metric compatible connection
the subspace $\mathbf{R}^{(3)}$ has the highest dimension,
so it will be easier to fit our curvature into this subspace.

Let $R$ be the Riemann curvature generated by a
metric compatible connection .
Then $R\in\mathbf{R}^{(3)}$ if and only if
\begin{eqnarray}
\label{torsionwaves1}
R^T&=&R,
\\
\label{einsteinequation4}
Ric&=&0,
\\
\label{torsionwaves2}
\varepsilon^{\kappa\lambda\mu\nu}R_{\kappa\lambda\mu\nu}&=&0,
\end{eqnarray}
where transposition is defined in
accordance with formula (\ref{appendixatransposition}).
Equation (\ref{einsteinequation4})
in the above system
can, of course,
be replaced by the pair of equations
${}^*\!R^*=-R$, $\mathcal{R}=0$.

\begin{definition}
We call a metric compatible solution of the system
(\ref{torsionwaves1})--(\ref{torsionwaves2})
a \emph{Weyl pseudoinstanton}.
\end{definition}

This terminology is motivated by the fact that such
Riemann curvatures are purely Weyl.

In the next two sections we construct explicitly two
families of Weyl pseudo\-instantons.

\section{The vacuum Einstein equation}
\label{einsteinequation}

In this section we look for Riemannian Weyl pseudo\-instantons,
that is, for Weyl pseudoinstantons with zero torsion.
In this case the connection is Levi-Civita and
equations (\ref{torsionwaves1}), (\ref{torsionwaves2})
are automatically satisfied. This leaves us with
equation (\ref{einsteinequation4})
which is the vacuum Einstein equation.

Thus, the vacuum Einstein equation is simply the
explicit description of a Riemannian Weyl pseudoinstanton.

\section{Torsion waves}
\label{torsionwaves}

In this section we look for non-Riemannian Weyl pseudo\-instantons,
that is, for Weyl pseudoinstantons with non-zero torsion.
Throughout the section we work in Minkowski space
which we define as a real 4-manifold with
global coordinate system $(x^0,x^1,x^2,x^3)$ and metric
\begin{equation}
\label{metric}
g_{\mu\nu}={\rm diag}(+1,-1,-1,-1)\,.
\end{equation}
Note that our definition of Minkowski space
specifies the manifold $M$ and the metric $g$,
but does not specify the connection $\Gamma$.

The construction we are about to carry out is, in a sense,
the opposite of what we did in the previous section:
in Section \ref{einsteinequation} we looked for
Weyl pseudoinstantons with non-trivial metric and zero torsion,
whereas now we will be looking for
Weyl pseudoinstantons with trivial metric (\ref{metric})
and non-zero torsion.
It is important to emphasize that the fact that the metric is constant
does not imply that curvature is zero because the connection
coefficients appearing in (\ref{riemanncurvature}) are not necessarily
Christoffel symbols.

Formulae
(\ref{contortionviagammaandchristoffel}),
(\ref{christoffel}), (\ref{metric})
imply ${K^\lambda}_{\mu\nu}={\Gamma^\lambda}_{\mu\nu}$,
so in the Minkowski setting
the system (\ref{torsionwaves1})--(\ref{torsionwaves2})
is a system of first order partial differential equations
for the unknown contortion.
There are two difficulties associated with this system.
Firstly, it is overdetermined:
the number of independent equations is $15+10+1$
whereas the number of unknowns
(independent components of the contortion tensor) is only 24.
Secondly, it has a quadratic nonlinearity
resulting from the commutator in
the formula for Riemann curvature
(last two terms in the RHS of
formula (\ref{riemanncurvature})).

The second difficulty is fundamental,
however it can be overcome by means of
the \emph{linearization ansatz} suggested in
\cite{kingandvassiliev,kiev}.
Namely, one seeks the unknown contortion in the form
${K^\lambda}_{\mu\nu}={\rm Re}(v_\mu{L^\lambda}_\nu)$
where $v$ is a complex-valued vector function
and $L\ne0$ is a constant complex antisymmetric
tensor satisfying $*L=\pm iL$. Then
the nonlinear system (\ref{torsionwaves1})--(\ref{torsionwaves2})
turns into a \emph{linear} system for the vector function $v$.
The coefficients of this linear system
of partial differential equations
depend on the tensor $L$ as a parameter,
and this parameter dependence is also linear.

It is interesting that the idea of seeking the unknown rank 3 tensor
in the form of a product of a vector and a rank 2 tensor
(``separation of indices'')
goes back to Lanczos, see formula (XI.1) in his paper \cite{lanczos1949}.
Unfortunately, Lanczos did not develop this idea.
He restricted his analysis to the following observation:
``Such solutions cannot be studied on the basis
of purely linear operators\ldots\ Hence they are outside the limits
of the present investigation.''

Explicit calculations \cite{kingandvassiliev,kiev} produce
a Weyl pseudo\-instanton which can be written down in the
following compact form. This metric compatible spacetime is
characterized by metric (\ref{metric}) and torsion
\begin{equation}
\label{torsion}
{T^\lambda}_{\mu\nu}
=\frac12{\rm Re}(u^\lambda(du)_{\mu\nu})
\end{equation}
where $u$  is
a non-trivial plane wave solution of the polarized Maxwell equation
\begin{equation}
\label{polarizedmaxwellequation}
*du=\pm idu.
\end{equation}
Here $d$ is the operator of exterior differentiation,
$u$ is a complex-valued vector function,
``plane wave'' means that
$u(x)=w\,e^{-ik\cdot x}$
where $w$ is a constant complex vector
and $k$ is a constant real vector,
and ``non-trivial'' means that $du\not\equiv0$.

Let us stress that the spacetime
(\ref{metric})--(\ref{polarizedmaxwellequation})
is a solution of the full nonlinear system
(\ref{torsionwaves1})--(\ref{torsionwaves2}),
and, consequently, a solution of the full nonlinear system
(\ref{eulerlagrangemetric}), (\ref{eulerlagrangeconnection}).

A detailed analysis of the solution
(\ref{metric})--(\ref{polarizedmaxwellequation})
carried out in \cite{kingandvassiliev,kiev} shows that it may be
interpreted as the neutrino. This interpretation
is based on the examination of the corresponding Riemann curvature
\begin{equation}
\label{curvature}
R_{\kappa\lambda\mu\nu}
={\rm Re}((du)_{\kappa\lambda}(du)_{\mu\nu}).
\end{equation}
Clearly, $(du)_{\kappa\lambda}(du)_{\mu\nu}$
is a factorized Weyl curvature, which according to
Lemma \ref{factorizationofcurvature}
makes it equivalent to a spinor
$\xi=\left(\begin{array}{c}\xi^1\\\xi^2\end{array}\right)$
or
$\eta=\left(\begin{array}{c}\eta_{\dot1}\\\eta_{\dot2}\end{array}\right)$.
It turns out that this spinor function satisfies
the appropriate half of Weyl's equation
$\gamma^\mu\partial_\mu\psi=0$, that is,
\[
\partial_0\xi+
(\sigma^1\partial_1+\sigma^2\partial_2+\sigma^3\partial_3)\xi=0
\]
or
\[
\partial_0\eta-
(\sigma^1\partial_1+\sigma^2\partial_2+\sigma^3\partial_3)\eta=0.
\]

\begin{remark}
It is known (see, for example, \cite{ryder}) that in a non-Riemannian
spacetime with metric compatible connection Weyl's equation
has an additional term with torsion. However, this additional term
involves only the axial component of torsion which in our case is zero.
See Appendix B.2 in \cite{hehlreview}
or Appendix \ref{appendixb} in our paper
for irreducible decomposition of torsion.
\end{remark}

In interpreting the solution
(\ref{metric})--(\ref{polarizedmaxwellequation})
as the neutrino we chose to deal with curvature rather than
torsion because curvature is an accepted physical observable.
If we also accept torsion as a
physical observable then the situation changes.
Given a plane wave solution $u$
of the polarized Maxwell equation
(\ref{polarizedmaxwellequation}) we can always add to it the gradient
of a scalar plane wave,
which changes torsion (\ref{torsion}) but does not change curvature
(\ref{curvature}).
Thus,
different torsions can generate the same
curvature, and having accepted torsion as a
physical observable we have to treat these solutions as different
particles. This might explain the subtle difference between
the electron, muon and tau neutrinos.

Recently there has been a series of publications
\cite{MR1679144,MR99d83082,MR2000j83067,MR1803749,MR1779651}
in which the authors constructed other types of torsion waves.
Our torsion waves (\ref{metric})--(\ref{polarizedmaxwellequation})
are fundamentally different from those in
\cite{MR1679144,MR99d83082,MR2000j83067,MR1803749,MR1779651}.
The differences are as follows.
\begin{itemize}
\item
The action in
\cite{MR1679144,MR99d83082,MR2000j83067,MR1803749,MR1779651}
is more general in that it
contains terms with torsion and nonmetricity.
However, the solutions found in these publications
become Riemannian for our action (\ref{action}).
\item
The metric in
\cite{MR1679144,MR99d83082,MR2000j83067,MR1803749,MR1779651}
is non-constant and
the connection is not metric compatible,
whereas our metric is constant
and connection is metric compatible.
\item
The torsion in
\cite{MR1679144,MR99d83082,MR2000j83067,MR1803749,MR1779651}
is purely trace,
whereas ours is purely tensor.
\item
Our torsion and Riemann curvature are monochromatic plane waves, i.e.,
\begin{eqnarray*}
T(x)&=&T'\cos(k\cdot x)+T''\sin(k\cdot x),
\\
R(x)&=&R'\cos(k\cdot x)+R''\sin(k\cdot x),
\end{eqnarray*}
where $T'$, $T''$, $R'$, and $R''$ are constant real tensors.
In \cite{MR1679144,MR99d83082,MR2000j83067,MR1803749,MR1779651}
torsion and curvature
are more complex.
\end{itemize}

The most important feature of
our torsion wave (\ref{metric})--(\ref{polarizedmaxwellequation})
is the fact that
\begin{equation}
\label{riemannianuniverse}
R\in\mathbf{R}^{(1)}\oplus\mathbf{R}^{(2)}\oplus\mathbf{R}^{(3)}.
\end{equation}
The RHS of formula (\ref{riemannianuniverse}) is the space of
Riemann curvatures
generated by Levi-Civita connections. Formula
(\ref{riemannianuniverse}) means that the curvature
generated by our torsion wave has all the symmetry properties
of the usual curvature from Riemannian geometry.
Therefore,
in observing such a torsion wave we might
not interpret it as torsion at all and believe that
we live in a Riemannian universe.

The torsion waves in
\cite{MR1679144,MR99d83082,MR2000j83067,MR1803749,MR1779651}
do not possess the property (\ref{riemannianuniverse}).

\section{The Ricci pseudoinstanton}
\label{riccipseudoinstanton}

Let $R$ be the Riemann curvature generated by a
metric compatible connection.
Then $R\in\mathbf{R}^{(1)}$ if and only if
equations (\ref{torsionwaves1}), (\ref{torsionwaves2}), and
\begin{eqnarray}
\label{ricci1}
\mathcal{R}&=&0,
\\
\label{ricci2}
\mathcal{W}&=&0
\end{eqnarray}
are satisfied.
The last three equations in the system
(\ref{torsionwaves1}), (\ref{torsionwaves2}),
(\ref{ricci1}), (\ref{ricci2})
can, of course, be replaced by the equation
${}^*\!R^*=R$.

\begin{definition}
\label{definitionofariccipseudoinstanton}
We call a metric compatible solution of the system
(\ref{torsionwaves1}), (\ref{torsionwaves2}),
(\ref{ricci1}), (\ref{ricci2})
a \emph{Ricci pseudoinstanton}.
\end{definition}

This terminology is motivated by the fact that such
Riemann curvatures
are completely determined by the trace free Ricci tensor.

We cannot produce torsion wave solutions of the system
(\ref{torsionwaves1}), (\ref{torsionwaves2}),
(\ref{ricci1}), (\ref{ricci2}). More precisely,
our linearization ansatz \cite{kingandvassiliev,kiev}
when applied to this system does not produce non-trivial
($R\not\equiv0$) solutions.
There are, however, Riemannian solutions.

\begin{definition}
\label{definitionofathompsonmanifold}
We call a Riemannian spacetime a \emph{Thompson space}
if its scalar and Weyl curvatures
are zero.
\end{definition}

Thompson noticed \cite{thompsonFeb1975} that
such spaces satisfy equation (\ref{eulerlagrangeconnection}).
Later Fairchild
addressed the question whether Thompson spaces satisfy
equation (\ref{eulerlagrangemetric}). He first thought
\cite{fairchild1976} that they do not,
but in a subsequent erratum \cite{fairchild1976erratum}
concluded that Thompson spaces do indeed satisfy
equation (\ref{eulerlagrangemetric}).
Thompson and Fairchild carried out their analysis for the
Yang--Mills case (\ref{specialcase}) but the result remains
true for arbitrary weights $c_j$ because Thompson spaces
are Ricci pseudo\-instantons.

The physical meaning of Thompson spaces is unclear.
It has been suggested by
Thompson \cite{thompsonFeb1975,thompsonAug1975},
Pavelle \cite{pavelleApr1975},
and Fairchild \cite{fairchild1976,fairchild1976erratum}
that these are nonphysical solutions.

It is worth noting that in appropriate local coordinates
the metric of a Thompson space can be written as
$g_{\mu\nu}=e^{2f}\,{\rm diag}(+1,-1,-1,-1)$ where
$f$ is a real scalar function satisfying
$\square f+\|{\rm grad}f\|^2=0$.
Here $\,\square:=\partial_\mu\partial^\mu\,$,
$\,({\rm grad}f)_\mu:=\partial_\mu f\,$,
$\,\|v\|^2:=v_\mu v^\mu\,$, and the
raising of indices is performed with respect
to the Minkowski metric (\ref{metric}).
Our problem
(\ref{eulerlagrangemetric}), (\ref{eulerlagrangeconnection})
is conformally invariant (see Remark \ref{conformal}),
therefore the natural thing to do is to
rescale the metric and view such a solution as a scalar field on a
manifold with Minkowski metric.

\section{Einstein spaces}
\label{einsteinmanifolds}

In this section we look for solutions of the system
(\ref{eulerlagrangemetric}), (\ref{eulerlagrangeconnection})
which are not pseudo\-instantons. As this is an
exceptionally difficult
mathematical problem we restrict our search to Riemannian spacetimes.

Lengthy but straightforward calculations give the
following explicit representation for equations
(\ref{eulerlagrangemetric}), (\ref{eulerlagrangeconnection}):
\begin{equation}
\label{einsteinmanifolds2}
(c_1+c_3)\mathcal{W}^{\kappa\lambda\mu\nu}\mathcal{R}ic_{\kappa\mu}
+\frac{c_1+c_2}6
\,\mathcal{R}\,\mathcal{R}ic^{\lambda\nu}=0,
\end{equation}
\begin{multline}
\label{einsteinmanifolds1}
(c_1+c_3)
(\nabla_\lambda\mathcal{R}ic_{\kappa\mu}
-\nabla_\kappa\mathcal{R}ic_{\lambda\mu})
\\
+\left(\frac{c_1}4+\frac{c_2}6+\frac{c_3}{12}\right)
(g_{\kappa\mu}\partial_\lambda\mathcal{R}
-g_{\lambda\mu}\partial_\kappa\mathcal{R})=0.
\end{multline}
Weyl curvature
has been excluded from equation (\ref{einsteinmanifolds1}) by
means of the Bianchi identity. Note that this ``trick''
does not work for a general affine connection,
nor does it work for a general metric compatible connection.

\begin{definition}
We call a Riemannian spacetime an \emph{Einstein space} if
its Ricci curvature and metric are related as
\begin{equation}
\label{einsteinmanifolds3}
Ric=\Lambda g
\end{equation}
where $\Lambda$ is some real ``cosmological'' constant.
\end{definition}

Alternatively, an Einstein space can be defined as a
Riemannian spacetime with
\begin{equation}
\label{einsteinmanifolds5}
\mathcal{R}ic=0.
\end{equation}
Formula (\ref{einsteinmanifolds5}) and the contracted Bianchi
identity imply
\begin{equation}
\label{einsteinmanifolds4}
\mathcal{R}=4\Lambda
\end{equation}
with some constant $\Lambda$.
The pair of conditions
(\ref{einsteinmanifolds5}), (\ref{einsteinmanifolds4})
is, of course, equivalent to condition (\ref{einsteinmanifolds3}).

Clearly, equations
(\ref{einsteinmanifolds5}), (\ref{einsteinmanifolds4})
imply equations
(\ref{einsteinmanifolds2}), (\ref{einsteinmanifolds1}),
so Einstein spaces are solutions of our problem
(\ref{eulerlagrangemetric}), (\ref{eulerlagrangeconnection}).
Thus, our model with arbitrary weights $c_j$ inherits
the main feature of the Yang--Mielke theory of gravity.

\begin{remark}
The fact that Einstein spaces are solutions of the system
(\ref{eulerlagrangemetric}), (\ref{eulerlagrangeconnection})
in the case of arbitrary weights $c_j$ was already
known to Buchdahl \cite{buchdahl}.
Buchdahl's review appears to have escaped the attention
of subsequent researchers in the subject area.
\end{remark}

\section{Discussion of the Riemannian case}
\label{riemannian}

We have found two families of Riemannian solutions to our problem
(\ref{eulerlagrangemetric}), (\ref{eulerlagrangeconnection}),
namely, Thompson and Einstein spaces.
Thompson spaces fit into our pseudo\-instanton scheme
whereas Einstein spaces do not (their Riemann curvature
has, in general, two non-trivial irreducible pieces).
It is natural to attempt to
explain why Einstein spaces are solutions without having to
write down explicitly the Euler--Lagrange equations.
The explanation is as follows.
In order to adapt the arguments from Section
\ref{proofofmaintheorem} to the case of an Einstein space
it is necessary and sufficient to show that
\begin{eqnarray}
\label{riemannian1}
(c_2-c_3)\,\partial(\|R^{(2)}\|^2)/\partial g&=&0,
\\
\label{riemannian2}
(c_2-c_3)\,\partial(\|R^{(2)}\|^2)/\partial\Gamma&=&0.
\end{eqnarray}
The case $c_2=c_3$ is trivial, so further on in this paragraph
we assume $c_2\ne c_3$.
The reason why equations
(\ref{riemannian1}), (\ref{riemannian2})
are satisfied for an Einstein space is that
the irreducible piece $R^{(2)}$ has a very simple
structure. We have $\|R^{(2)}\|^2=-\frac16\int\mathcal{R}^2$,
and elementary calculations show that
the system (\ref{riemannian1}), (\ref{riemannian2})
is equivalent to
\begin{eqnarray}
\label{riemannian3}
\mathcal{R}\,\mathcal{R}ic&=&0,
\\
\label{riemannian4}
\partial\mathcal{R}&=&0.
\end{eqnarray}
These equations are clearly satisfied under the conditions
(\ref{einsteinmanifolds5}), (\ref{einsteinmanifolds4}).

In establishing that Thompson spaces are solutions of
the system
(\ref{eulerlagrangemetric}), (\ref{eulerlagrangeconnection})
we relied on our general pseudoinstanton construction,
without analyzing the actual Euler--Lagrange equations
which in the Riemannian case have the explicit representation
(\ref{einsteinmanifolds2}), (\ref{einsteinmanifolds1}).
It may be worrying that the inspection of
equation (\ref{einsteinmanifolds1}) does not immediately confirm
that for a Thompson space this equation is satisfied.
These fears are laid to rest if one rewrites
equation (\ref{einsteinmanifolds1})
in equivalent form excluding
the trace free Ricci curvature by means of the identity
\begin{equation}
\label{specialidentity}
\nabla_\lambda\mathcal{R}ic_{\kappa\mu}
-\nabla_\kappa\mathcal{R}ic_{\lambda\mu}
=-\frac1{12}(g_{\kappa\mu}\partial_\lambda\mathcal{R}
-g_{\lambda\mu}\partial_\kappa\mathcal{R})
+2\nabla_\nu\mathcal{W}_{\kappa\lambda\mu}{}^\nu
\end{equation}
(consequence of the Bianchi identity).
This turns equation (\ref{einsteinmanifolds1}) into
\[
(c_1+c_3)
\nabla_\nu\mathcal{W}_{\kappa\lambda\mu}{}^\nu
+\frac{c_1+c_2}{12}\,
(g_{\kappa\mu}\partial_\lambda\mathcal{R}
-g_{\lambda\mu}\partial_\kappa\mathcal{R})=0
\]
which is clearly satisfied under the conditions
(\ref{ricci1}), (\ref{ricci2}).

An interesting feature of the system
(\ref{einsteinmanifolds2}), (\ref{einsteinmanifolds1})
is that it does not contain the parameters $c_1$, $c_2$, $c_3$
separately, only their combinations $c_1+c_2$ and $c_1+c_3$.
This warrants an explanation which goes as follows.
We know (see Section \ref{proofofmaintheorem})
that for spacetimes with metric compatible connections
the expression
\[
\|R+{}^*\!R^*\|^2-\|R-{}^*\!R^*\|^2
\]
is a topological invariant.
Therefore,
\[
\delta(\|R+{}^*\!R^*\|^2)
=\delta(\|R-{}^*\!R^*\|^2).
\]
If we start our variation from a Riemannian spacetime
then the latter formula becomes
\begin{equation}
\label{riemannian5}
\delta(\|R^{(1)}\|^2)=
\delta(\|R^{(2)}\|^2)+\delta(\|R^{(3)}\|^2).
\end{equation}
Formula (\ref{riemannian5}) was written under the assumption
that variation preserves metric compatibility, however
arguments presented in the first paragraph of
Section \ref{proofofmaintheorem} show that it remains
true for arbitrary variations.
Formulae (\ref{action}), (\ref{riemannian5}) imply
\[
\delta S=
(c_1+c_2)\,\delta(\|R^{(2)}\|^2)+(c_1+c_3)\,\delta(\|R^{(3)}\|^2)
\]
which explains why the resulting
Euler--Lagrange equations contain only the combinations
of weights $c_1+c_2$ and $c_1+c_3$.

Finally, let us give a simple characterization of
Thompson and Einstein spaces.
It is easy to see that these spaces are of opposite double duality:
we have
\begin{eqnarray}
\label{doubledualityplus}
{}^*\!R^*&=&+R\,,
\\
\label{doubledualityminus}
{}^*\!R^*&=&-R
\end{eqnarray}
for Thompson and Einstein spaces respectively.
Moreover, a Riemannian spacetime is a Thompson space
if and only if it satisfies condition (\ref{doubledualityplus}),
and an Einstein space
if and only if it satisfies condition (\ref{doubledualityminus}).

\section{Uniqueness}
\label{uniqueness}

We have constructed in total
three families of solutions to our problem
(\ref{eulerlagrangemetric}), (\ref{eulerlagrangeconnection}):
torsion waves in Minkowski space (Section \ref{torsionwaves}),
Thompson spaces (Section \ref{riccipseudoinstanton}),
and Einstein spaces (Section \ref{einsteinmanifolds}).
The question we are about to address is whether these
three families are \emph{all} the solutions of the problem
(\ref{eulerlagrangemetric}), (\ref{eulerlagrangeconnection})
within suitable classes of solutions.
As questions of uniqueness in metric-affine field theory
are notoriously difficult we will be forced to argue
mostly at the level of conjectures.

\begin{conjecture}
\label{conjecture1}
For generic weights $\,c_j\,$
torsion waves constructed in Section \ref{torsionwaves}
are the only solutions of our problem
(\ref{eulerlagrangemetric}), (\ref{eulerlagrangeconnection})
among connections of the type
\[
\Gamma(x)=\Gamma'\cos(k\cdot x)+\Gamma''\sin(k\cdot x),\quad
k\ne0,\quad R(x)\not\equiv0
\]
in Minkowski space.
\end{conjecture}

Conjecture \ref{conjecture1} is motivated by the fact that in
the Minkowski setting (\ref{metric}) the system
(\ref{eulerlagrangemetric}), (\ref{eulerlagrangeconnection})
is heavily overdetermined:
it is a system of $9+64$ equations with only 64 unknowns.

The construction carried out in Section \ref{torsionwaves}
is effectively based on the use of hidden symmetries of our problem,
and there is no obvious way of generalizing it unless
there are some additional symmetries due to a special choice
of weights $c_j$.
An example of such a special choice is the Yang--Mills case
(\ref{specialcase}). Theorem 1 from \cite{kingandvassiliev}
establishes that in this case the problem
(\ref{eulerlagrangemetric}), (\ref{eulerlagrangeconnection})
has a wider family of torsion wave solutions than those
described in Section \ref{torsionwaves}.
Thus, the Yang--Mills case (\ref{specialcase}) is not generic in the
sense of Conjecture \ref{conjecture1}.

The fact that in the Yang--Mills case (\ref{specialcase})
the problem
(\ref{eulerlagrangemetric}), (\ref{eulerlagrangeconnection})
has too many torsion wave solutions leads to serious difficulties.
In \cite{kingandvassiliev} we were unable to attribute a physical
interpretation to all these solutions, and, in order to reduce the
number of solutions, were forced to introduce the
vacuum Einstein equation (\ref{einsteinequation4}) as an
additional equation in our model. In the current paper this
difficulty has been overcome by switching from the Yang--Mills
action $\|R\|^2$ to the action (\ref{action}) with arbitrary
weights $c_j$. In this case
the only general tool at our disposal
is the pseudo\-instanton construction which naturally leads
to the vacuum Einstein equation (\ref{einsteinequation4}).

\begin{conjecture}
\label{conjecture2}
For generic weights $\,c_j\,$
Thompson and Einstein spaces are the only Riemannian solutions of
our problem
(\ref{eulerlagrangemetric}), (\ref{eulerlagrangeconnection}).
\end{conjecture}

Conjecture \ref{conjecture2} is motivated by the following arguments.
Suppose
\begin{equation}
\label{uniqueness1}
c_1+c_2\ne0,\quad c_1+c_3\ne0.
\end{equation}
Then the system
(\ref{einsteinmanifolds2}), (\ref{einsteinmanifolds1})
is equivalent to the following system:
equation (\ref{einsteinmanifolds4}) with some constant $\Lambda$ and
equations
\begin{equation}
\label{uniqueness3}
\mathcal{W}^{\kappa\lambda\mu\nu}\mathcal{R}ic_{\kappa\mu}
+c\,\mathcal{R}\,\mathcal{R}ic^{\lambda\nu}
=0,
\end{equation}
\begin{equation}
\label{uniqueness2}
\nabla_\lambda\mathcal{R}ic_{\kappa\mu}
-\nabla_\kappa\mathcal{R}ic_{\lambda\mu}=0,
\end{equation}
where
\begin{equation}
\label{parameterc}
c:=\frac{c_1+c_2}{6(c_1+c_3)}\ne0
\end{equation}
is a dimensionless parameter.
(Of course, by virtue of the identity (\ref{specialidentity}) the pair
of equations (\ref{einsteinmanifolds4}) and (\ref{uniqueness2})
is equivalent to the pair
of equations (\ref{einsteinmanifolds4}) and
$\nabla_\nu\mathcal{W}_{\kappa\lambda\mu}{}^\nu=0$.)
The constant $\Lambda$ in (\ref{einsteinmanifolds4})
is either 0 or it scales to $\pm1$, so the system
(\ref{einsteinmanifolds4}),
(\ref{uniqueness3}), (\ref{uniqueness2})
effectively contains only one free parameter, $c$.
The number of independent equations in
(\ref{einsteinmanifolds4}),
(\ref{uniqueness3}), (\ref{uniqueness2})
is $1+9+16$ whereas the number of unknowns
(independent components of the metric tensor) is only 10.
It is hard to imagine how this overdetermined
system can have solutions without the symmetry
(\ref{doubledualityplus}) or (\ref{doubledualityminus}),
except for some special values of the parameter $c$.

The search for special values of $c$ for which our problem has
more Riemannian solutions than expected is
similar to the Cosserat problem in the theory of elasticity,
which is the study of the
elasticity operator with Poisson's ratio
treated as the spectral parameter;
see \cite{MR93i35099} and the extensive bibliographic list therein
for details.
In our case the parameter (\ref{parameterc})
plays the role of Poisson's ratio.

As an illustration let us consider the
$2+2$ decomposable case
when our 4-manifold is the product of two Riemannian 2-manifolds,
see \cite{thompsonAug1975} or \cite{MR2001a83023} for details.
Straightforward calculations establish
the following result within this class of solutions:
\begin{itemize}
\item
if $c\ne-\frac13$ then Thompson and
Einstein spaces are the only solutions of
the problem
(\ref{eulerlagrangemetric}), (\ref{eulerlagrangeconnection}),
whereas
\item
if $c=-\frac13$ then Thompson and
Einstein spaces are not the only solutions of
the problem
(\ref{eulerlagrangemetric}), (\ref{eulerlagrangeconnection}).
\end{itemize}

We see that the case $\,c=-\frac13\,$ is not generic in the sense of
Conjecture \ref{conjecture2}.
It is interesting that
condition (\ref{specialcase}) implies condition (\ref{uniqueness1})
as well as $\,c\ne-\frac13\,$, so it may be that
the Yang--Mills case is generic in the sense of
Conjecture \ref{conjecture2}.

One should have in mind that
the problem of uniqueness is very delicate
even in the Yang--Mills case (\ref{specialcase})
and even within the class of Riemannian solutions.
Fairchild's attempt \cite{fairchild1976}
at establishing uniqueness for the problem
(\ref{eulerlagrangemetric}), (\ref{eulerlagrangeconnection})
was unsuccessful:
the result and its proof were incorrect and the author had
to publish an erratum \cite{fairchild1976erratum}.

\section{The Bach action}
\label{bachaction}

In this section we consider the case when the weights
appearing in formula (\ref{action}) are
\begin{equation}
\label{bach1}
c_3=1\quad{\rm and}\quad c_j=0,\quad\forall j\ne3.
\end{equation}
In this case our action becomes
\begin{equation}
\label{bach2}
S=\|\mathcal{W}\|^2,
\end{equation}
where $\mathcal{W}=R^{(3)}$ is the Weyl curvature.
The action (\ref{bach2}) is called the \emph{Bach action}.
If one assumes the connection to be Levi-Civita and varies
(\ref{bach2}) with respect to the metric then
the resulting Euler--Lagrange equation is
the classical Bach equation \cite{bach};
see also \cite{MR2001a83023} for an account of the
modern state of the subject.
Our approach is to vary the metric and the connection independently,
which leads to the metric-affine version
(\ref{eulerlagrangemetric}), (\ref{eulerlagrangeconnection})
of the Bach equation.

All the general results obtained in this paper apply to the
Bach case (\ref{bach1}). However, the peculiarity of this case
is that any spacetime with zero Weyl curvature is a solution
of the metric-affine Bach problem
(\ref{eulerlagrangemetric}), (\ref{eulerlagrangeconnection});
here we do not have to make any assumptions concerning the other
irreducible pieces of curvature as we did in
Section \ref{riccipseudoinstanton}.
Accordingly, the analysis of Riemannian solutions
has to be modified (note that in the Bach case
the first condition (\ref{uniqueness1}) is not satisfied).
Equations
(\ref{einsteinmanifolds2}), (\ref{einsteinmanifolds1})
are now equivalent to
\begin{eqnarray}
\label{bach3}
\mathcal{W}^{\kappa\lambda\mu\nu}\mathcal{R}ic_{\kappa\mu}
=0,
\\
\label{bach4}
\nabla_\lambda\mathcal{R}ic_{\kappa\mu}
-\nabla_\kappa\mathcal{R}ic_{\lambda\mu}
+\frac1{12}(g_{\kappa\mu}\partial_\lambda\mathcal{R}
-g_{\lambda\mu}\partial_\kappa\mathcal{R})
\equiv2\nabla_\nu\mathcal{W}_{\kappa\lambda\mu}{}^\nu
=0,
\end{eqnarray}
where we made use of the identity (\ref{specialidentity}).
Equation (\ref{bach4}) does not imply
that the scalar curvature is constant, so we lose the equation
(\ref{einsteinmanifolds4}) which we previously
derived under the assumption $c_1+c_2\ne0$.
Nevertheless, the system
(\ref{bach3}), (\ref{bach4})
remains heavily overdetermined.
It is natural to state the following

\begin{conjecture}
\label{conjecture3}
In the Bach case (\ref{bach1}) the only
Riemannian solutions of our problem
(\ref{eulerlagrangemetric}), (\ref{eulerlagrangeconnection})
are conformally flat spaces and Einstein spaces.
\end{conjecture}

Though providing a rigorous proof of Conjecture \ref{conjecture3}
might be very difficult, one
can easily check that it is true
within the class of $2+2$ decomposable solutions.

Recall that the classical Bach equation has far more
solutions than stated in Conjecture~\ref{conjecture3}.
Spacetimes which are conformally related to Einstein spaces
are solutions, as are some non-trivial spacetimes found in
\cite{schmidt1} and \cite{MR2001a83023}.
Note that the paper \cite{MR2001a83023}
is based on the $2+2$ decomposition.

There is, of course, a fundamental difference in the role played
by the Bach action in the classical (Riemannian)
and metric-affine settings.
In the classical setting the Bach action is very special in that
it is constructed from the only conformally invariant
irreducible piece of curvature,
whereas in the metric-affine setting
it loses its special status because all the
irreducible pieces of curvature become conformally invariant;
see also Remark \ref{conformal}.

\section*{Acknowledgments}

The author is grateful to
D.~V.~Alekseevsky, F.~E.~Burstall, and A.~D.~King
for helpful advice, and to G.~Friesecke for
translating excerpts from \cite{weylquadraticaction}.

\appendix

\section{Irreducible decomposition of curvature}
\label{appendixa}

We give below an overview of
Appendix B.4 from \cite{hehlreview}, and present
the results in a form suitable to our needs.

A Riemann curvature generated by a
general affine connection has only one (anti)symmetry, namely,
\begin{equation}
\label{appendixa1}
R_{\kappa\lambda\mu\nu}=-R_{\kappa\lambda\nu\mu}\,.
\end{equation}
For a fixed $x\in M$
we denote by $\mathbf{R}$ the 96-dimensional vector space
of real rank 4 tensors satisfying condition (\ref{appendixa1}),
and we equip $\mathbf{R}$ with the natural
indefinite inner product
\begin{equation}
\label{innerproductoncurvaturesatgivenpoint}
(R,Q)_x:=R^\kappa{}_{\lambda\mu\nu}\,
Q^\lambda{}_\kappa{}^{\mu\nu}\,.
\end{equation}

We have the orthogonal decomposition
$\mathbf{R}=\mathbf{R}^+\oplus\mathbf{R}^-$
where
\[
\mathbf{R}^\pm=\{R\in\mathbf{R}|
R_{\kappa\lambda\mu\nu}=\pm R_{\lambda\kappa\mu\nu}\}.
\]
It is easy to see that
${\rm dim}\,\mathbf{R}^+=60$ and ${\rm dim}\,\mathbf{R}^-=36$.

The subspaces $\mathbf{R}^+$ and $\mathbf{R}^-$ decompose
further into five and six irreducible subspaces respectively.
We are mostly interested in $\mathbf{R}^-$ as
this is the vector space of curvatures generated by metric
compatible connections, so what follows is a
description of the irreducible subspaces of $\mathbf{R}^-$.

Put
\begin{eqnarray}
\label{appendixatransposition}
(R^T)_{\kappa\lambda\mu\nu}&:=&R_{\mu\nu\kappa\lambda}\,,
\\
\label{appendixa2}
({}^*\!R)_{\kappa\lambda\mu\nu}&:=&
\frac12\,\sqrt{|\det g|}
\ \varepsilon^{\kappa'\lambda'}{}_{\kappa\lambda}\,
R_{\kappa'\lambda'\mu\nu}\,,
\\
\label{appendixa3}
(R^*)_{\kappa\lambda\mu\nu}&:=&
\frac12\,\sqrt{|\det g|}
\ R_{\kappa\lambda\mu'\nu'}\,
\varepsilon^{\mu'\nu'}{}_{\mu\nu}\,.
\end{eqnarray}
The maps
\begin{eqnarray}
\label{endomorphism1}
R&\to&R^T,
\\
\label{endomorphism2}
R&\to&{}^*\!R\,,
\\
\label{endomorphism3}
R&\to&R^*
\end{eqnarray}
are endomorphisms in $\mathbf{R}^-$.
We call them \emph{transposition},
\emph{left Hodge star} and \emph{right Hodge star}
respectively. The left Hodge star acts on the internal (Lie algebra)
indices of curvature, whereas the right Hodge star acts on
the external ones and is the Hodge star used in abstract
Yang--Mills theory.

The eigenvalues of the map (\ref{endomorphism1}) are $\pm1$,
whereas the maps
(\ref{endomorphism2}) and (\ref{endomorphism3}) have
no eigenvalues at all (as we are working in the real
setting $\pm i$ are not eigenvalues).
This impediment is overcome
by working with the map
\begin{equation}
\label{endomorphism4}
R\to{}^*\!R^*
\end{equation}
rather than with the maps
(\ref{endomorphism2}) and (\ref{endomorphism3}) separately.
Here
\begin{equation}
\label{doubleduality}
{}^*\!R^*:=({}^*\!R)^*={}^*\!(R^*),
\end{equation}
and the order of operations
does not matter because the maps
(\ref{endomorphism2}) and (\ref{endomorphism3})
commute. We call
the endomorphism (\ref{endomorphism4}) the \emph{double duality}
map. Its eigenvalues are $\pm1$.

Clearly, the maps (\ref{endomorphism1}) and (\ref{endomorphism4})
commute and square to the identity, so
$\mathbf{R}^-=\underset{a,b=\pm}{\oplus}\mathbf{R}^-_{ab}$
where
\[
\mathbf{R}^-_{ab}=\{R\in\mathbf{R}^-|R^T=aR,{}^*\!R{}^*=bR\}.
\]
The maps
(\ref{endomorphism1}) and (\ref{endomorphism4})
are formally
self-adjoint with respect to the inner product
(\ref{innerproductoncurvaturesatgivenpoint})
so the subspaces
$\mathbf{R}^-_{++}$,
$\mathbf{R}^-_{+-}$,
$\mathbf{R}^-_{-+}$, and
$\mathbf{R}^-_{--}$
are mutually orthogonal. Their dimensions turn out to be
9, 12, 9, and 6 respectively.

For a Riemann curvature
$R\in\mathbf{R}^-_{++}$ the corresponding Ricci curvature
is symmetric trace free, and it completely determines $R$ itself
according to the formula
\[
R_{\kappa\lambda\mu\nu}=
\frac12
(g_{\kappa\mu}Ric_{\lambda\nu}
-g_{\lambda\mu}Ric_{\kappa\nu}
-g_{\kappa\nu}Ric_{\lambda\mu}
+g_{\lambda\nu}Ric_{\kappa\mu}).
\]
For a Riemann curvature
$R\in\mathbf{R}^-_{--}$ the corresponding Ricci curvature
is antisymmetric, and it completely determines $R$ itself
according to the same formula.
The subspace $\mathbf{R}^-_{-+}$ is the image
of $\mathbf{R}^-_{++}$ under either of the maps
(\ref{endomorphism2}) or (\ref{endomorphism3}).
We see that each of the subspaces
$\mathbf{R}^-_{++}$,
$\mathbf{R}^-_{-+}$,
$\mathbf{R}^-_{--}$
is equivalent to a space of real rank 2 tensors,
either symmetric trace free or antisymmetric. Therefore
these three subspaces are irreducible.

The only subspace which decomposes further is $\mathbf{R}^-_{+-}$\,:
\[
\mathbf{R}^-_{+-}=
\mathbf{R}_{\rm scalar}
\oplus
\mathbf{R}_{\rm Weyl}
\oplus
\mathbf{R}_{\rm pseudoscalar}\,.
\]
Here $\mathbf{R}_{\rm scalar}$ and
$\mathbf{R}_{\rm pseudoscalar}$ are the 1-dimensional
spaces of real Riemann curvatures $R_{\kappa\lambda\mu\nu}$
proportional to
$g_{\kappa\mu}g_{\lambda\nu}-g_{\lambda\mu}g_{\kappa\nu}$
and $\varepsilon_{\kappa\lambda\mu\nu}$ respectively,
and $\mathbf{R}_{\rm Weyl}$ is their 10-dimensional
orthogonal complement.

The decomposition described above assumes
curvature to be real and metric to be Lorentzian.
If curvature is complex or if $\det g>0$ then
the decomposition in somewhat different.
In particular, the subspaces
$\mathbf{R}_{\rm Weyl}$ and $\mathbf{R}^-_{--}$
decompose further into eigenspaces of the Hodge star
(left or right).

In order to simplify notation
in the main text we will denote the subspaces
\[
\mathbf{R}^-_{++}\,,\quad
\mathbf{R}_{\rm scalar}\,,\quad
\mathbf{R}_{\rm Weyl}\,,\quad
\mathbf{R}_{\rm pseudoscalar}\,,\quad
\mathbf{R}^-_{-+}\,,\quad
\mathbf{R}^-_{--}
\]
by $\mathbf{R}^{(j)}$, $j=1,\ldots,6$, respectively,
and the five subspaces of $\mathbf{R}^+$ by
$\mathbf{R}^{(j)}$, $j=7,\ldots,11$.

Thus, at each point $x\in M$
the vector space of Riemann curvatures decomposes as
$\mathbf{R}=\mathbf{R}^{(1)}\oplus\ldots\oplus\mathbf{R}^{(11)}$.
Consequently,
a Riemann curvature $R$ can be uniquely written as
$
R=R^{(1)}+\ldots+R^{(11)}
$
where $R^{(j)}\in\mathbf{R}^{(j)}$, $j=1,\ldots,11$,
are its irreducible pieces.
The explicit formulae for the first six pieces are
\begin{eqnarray*}
{R^{(1)}}_{\kappa\lambda\mu\nu}&=&\frac12
(g_{\kappa\mu}\overline{\mathcal{R}ic}_{\lambda\nu}
-g_{\lambda\mu}\overline{\mathcal{R}ic}_{\kappa\nu}
-g_{\kappa\nu}\overline{\mathcal{R}ic}_{\lambda\mu}
+g_{\lambda\nu}\overline{\mathcal{R}ic}_{\kappa\mu}),
\\
{R^{(2)}}_{\kappa\lambda\mu\nu}&=&\frac1{12}
(g_{\kappa\mu}g_{\lambda\nu}-g_{\lambda\mu}g_{\kappa\nu})
\mathcal{R}\,,
\\
R^{(3)}&=&\overline{R}-R^{(1)}-R^{(2)}-R^{(4)},
\\
{R^{(4)}}_{\kappa\lambda\mu\nu}&=&-\frac1{24}
\sqrt{|\det g|}\ \varepsilon_{\kappa\lambda\mu\nu}
\check{\mathcal{R}}\,,
\\
R^{(5)}&=&\widehat{R}-R^{(6)},
\\
{R^{(6)}}_{\kappa\lambda\mu\nu}&=&\frac12
(g_{\kappa\mu}\widehat{Ric}_{\lambda\nu}
-g_{\lambda\mu}\widehat{Ric}_{\kappa\nu}
-g_{\kappa\nu}\widehat{Ric}_{\lambda\mu}
+g_{\lambda\nu}\widehat{Ric}_{\kappa\mu}),
\end{eqnarray*}
where
\begin{eqnarray*}
&\overline{R}_{\kappa\lambda\mu\nu}=\frac14
(R_{\kappa\lambda\mu\nu}-R_{\lambda\kappa\mu\nu}
+R_{\mu\nu\kappa\lambda}-R_{\nu\mu\kappa\lambda})\,,
\\
&\widehat{R}_{\kappa\lambda\mu\nu}=\frac14
(R_{\kappa\lambda\mu\nu}-R_{\lambda\kappa\mu\nu}
-R_{\mu\nu\kappa\lambda}+R_{\nu\mu\kappa\lambda})\,,
\\
&\overline{Ric}_{\lambda\nu}=
\overline{R}\,{}^\kappa{}_{\lambda\kappa\nu}\,,
\qquad
\mathcal{R}=\overline{Ric}\,{}^\lambda{}_\lambda
=R^{\kappa\lambda}{}_{\kappa\lambda}\,,
\qquad
\overline{\mathcal{R}ic}_{\lambda\nu}=
\overline{Ric}_{\lambda\nu}-\frac14g_{\lambda\nu}\mathcal{R}\,,
\\
&\widehat{Ric}_{\lambda\nu}=\widehat{R}{}^\kappa{}_{\lambda\kappa\nu}\,,
\\
&\check{\mathcal{R}}=
\sqrt{|\det g|}\ \varepsilon^{\kappa\lambda\mu\nu}
\overline{R}_{\kappa\lambda\mu\nu}=
\sqrt{|\det g|}\ \varepsilon^{\kappa\lambda\mu\nu}
R_{\kappa\lambda\mu\nu}\,.
\end{eqnarray*}

Of course, in the Riemannian case curvature
has only three irreducible pieces, namely,
$R^{(1)}$, $R^{(2)}$, and $R^{(3)}$.

\section{Irreducible decomposition of torsion}
\label{appendixb}

According to Appendix B.2 from \cite{hehlreview} the irreducible
pieces of torsion are
\begin{eqnarray}
\label{appendixb1}
T^{(1)}&=&T-T^{(2)}-T^{(3)},
\\
\label{appendixb2}
{T^{(2)}}_{\lambda\mu\nu}&=&g_{\lambda\mu}v_\nu-g_{\lambda\nu}v_\mu\,,
\\
\label{appendixb3}
T^{(3)}&=&*w,
\end{eqnarray}
where
\begin{equation}
\label{appendixbvectorsviatorsion}
v_\nu=\frac13\,T^\lambda{}_{\lambda\nu}\,,
\qquad
w_\nu=\frac16\,\sqrt{|\det g|}\ T^{\kappa\lambda\mu}
\,\varepsilon_{\kappa\lambda\mu\nu}\,.
\end{equation}
The pieces $T^{(1)}$, $T^{(2)}$ and $T^{(3)}$
are called \emph{tensor torsion},
\emph{trace torsion}
and \emph{axial torsion} respectively.

We define the action of the Hodge star on torsions as
\begin{equation}
\label{appendixb4}
(*T)_{\lambda\mu\nu}:=
\frac12\,\sqrt{|\det g|}
\ T_{\lambda\mu'\nu'}\,
\varepsilon^{\mu'\nu'}{}_{\mu\nu}\,.
\end{equation}
The Hodge star maps tensor torsions to tensor torsions,
trace to axial, and axial to trace:
\begin{eqnarray}
\label{appendixb5}
(*T)^{(1)}&=&*(T^{(1)}),
\\
\label{appendixb6}
(*T)^{(2)}{}_{\lambda\mu\nu}&=&g_{\lambda\mu}w_\nu-g_{\lambda\nu}w_\mu,
\\
\label{appendixb7}
(*T)^{(3)}&=&-*v.
\end{eqnarray}
Note that the $*$ appearing in the RHS's of
formulae (\ref{appendixb3}) and (\ref{appendixb7})
is the standard Hodge star (\ref{hodgestar})
which should not be confused with
the Hodge star on torsions (\ref{appendixb4}).

The decomposition described above assumes
torsion to be real and metric to be Lorentzian.
If torsion is complex or if $\det g>0$ then
the subspace of tensor torsions
decomposes further into eigenspaces of the Hodge star.

Substituting formulae (\ref{appendixb1})--(\ref{appendixb3})
into formula (\ref{contortionviatorsion}),
and formula (\ref{torsionviacontortion})
into formulae (\ref{appendixbvectorsviatorsion})
we obtain the irreducible decomposition of contortion:
\begin{eqnarray}
\label{appendixb8}
K^{(1)}&=&K-K^{(2)}-K^{(3)},
\\
\label{appendixb9}
{K^{(2)}}_{\lambda\mu\nu}&=&g_{\lambda\mu}v_\nu-g_{\nu\mu}v_\lambda,
\\
\label{appendixb10}
K^{(3)}&=&\frac12*w,
\end{eqnarray}
where
\begin{equation}
\label{appendixbvectorsviacontortion}
v_\nu=\frac13\,K^\lambda{}_{\lambda\nu},
\qquad
w_\nu=\frac13\,\sqrt{|\det g|}\ K^{\kappa\lambda\mu}
\,\varepsilon_{\kappa\lambda\mu\nu}.
\end{equation}

The irreducible pieces of torsion
(\ref{appendixb1})--(\ref{appendixb3})
and contortion (\ref{appendixb8})--(\ref{appendixb10})
are related as
\[
{T^{(j)}}_{\lambda\mu\nu}={K^{(j)}}_{\mu\lambda\nu},
\quad j=1,2,
\qquad
{T^{(3)}}_{\lambda\mu\nu}=2{K^{(3)}}_{\lambda\mu\nu}
\]
(note the order of indices).

\section{Spinor representation of
Weyl curvature}
\label{appendixc}

Throughout this appendix we work in Minkowski space,
see (\ref{metric}). We follow Section 17 of \cite{LL4}
in our spinor notation, and we use the Latin letters
$a$, $b$, $c$, $d$ for spinor indices;
these run through the values $1$, $2$.
The Pauli and Dirac matrices are
\[
\sigma^1=\left(\begin{array}{cc}0&1\\1&0\end{array}\right),\qquad
\sigma^2=\left(\begin{array}{cc}0&-i\\ i&0\end{array}\right),\qquad
\sigma^3=\left(\begin{array}{cc}1&0\\0&-1\end{array}\right),
\]
\[
\gamma^0=\left(\begin{array}{cc}0&-I\\-I&0
\end{array}\right),\qquad
\gamma^j=\left(\begin{array}{cc}0&\sigma^j\\-\sigma^j&0
\end{array}\right),\quad j=1,2,3.
\]
We write rank 1 bispinors as columns
\[
\psi=\left(\begin{array}{c}
\xi^1\\\xi^2\\\eta_{\dot1}\\\eta_{\dot2}
\end{array}\right).
\]
The Dirac conjugate of $\psi$ is $\gamma^0\overline\psi$,
with the ``overline'' standing for complex conjugation.

Given a pair of tensor indices $\kappa$, $\lambda$
let us consider the matrix
$\gamma^0\gamma^2\gamma^\kappa\gamma^\lambda$
and write it in block form
\begin{equation}
\label{blockmatrix}
\gamma^0\gamma^2\gamma^\kappa\gamma^\lambda
=\left(
\begin{array}{cc}
(\gamma^0\gamma^2\gamma^\kappa\gamma^\lambda)_{ab}&
(\gamma^0\gamma^2\gamma^\kappa\gamma^\lambda)_a{}^{\dot b}\\
(\gamma^0\gamma^2\gamma^\kappa\gamma^\lambda)^{\dot a}{}_b&
(\gamma^0\gamma^2\gamma^\kappa\gamma^\lambda)^{\dot a\dot b}
\end{array}
\right).
\end{equation}
The diagonal blocks in the RHS of formula (\ref{blockmatrix}) are
symmetric for $\kappa\ne\lambda$ and
antisymmetric for $\kappa=\lambda$.
The off-diagonal blocks are
zero for all $\kappa$, $\lambda$.

It is easy to see that the matrix $\gamma^0\gamma^2$
represents a Lorentz invariant
linear map from the complex vector space
of rank 1 bispinors
to the complex vector space
of conjugate rank 1 bispinors.
Consequently, for an arbitrary rank 2 tensor $Q$
the matrix
$\gamma^0\gamma^2\gamma^\kappa\gamma^\lambda Q_{\kappa\lambda}$
has the same mapping property.
This explains the choice of spinor indices in
formula (\ref{blockmatrix}).

Let us recall the spinor
representation of an antisymmetric rank 2 tensor,
see Section 19 of \cite{LL4}
and Section 7 of \cite{kingandvassiliev} for details.
A complex antisymmetric rank 2 tensor $F$
is equivalent to a symmetric rank 2 bispinor
\begin{equation}
\label{appendixcaux}
\left(\begin{array}{c}\phi^{ab}\\
\chi_{\dot a\dot b}\end{array}\right),
\end{equation}
the relationship between the two being
\begin{equation}
\label{appendixc1}
F^{\kappa\lambda}=
(\gamma^0\gamma^2\gamma^\kappa\gamma^\lambda)_{ab}
\,\phi^{ab}
+
(\gamma^0\gamma^2\gamma^\kappa\gamma^\lambda)^{\dot a\dot b}
\,\chi_{\dot a\dot b}\,.
\end{equation}
Note that $*F=iF$ if and only if $\chi=0$,
and $*F=-iF$ if and only if $\phi=0$.

We say that a complex rank 4 tensor $R$ is a Weyl curvature if it
satisfies
\begin{eqnarray}
\label{appendixc2}
R_{\kappa\lambda\mu\nu}
=-R_{\lambda\kappa\mu\nu}
&=&-R_{\kappa\lambda\nu\mu}
=R_{\mu\nu\kappa\lambda}\,,
\\
\label{appendixc3}
{}^*\!R^*&=&-R\,,
\\
\label{appendixc4}
R^{\kappa\lambda}{}_{\kappa\lambda}&=&0,
\\
\label{appendixc5}
\varepsilon^{\kappa\lambda\mu\nu}R_{\kappa\lambda\mu\nu}&=&0.
\end{eqnarray}
Formulae (\ref{appendixcaux}),
(\ref{appendixc1}) allow us to give a spinor
representation of Weyl curvature. Namely,
a complex Weyl curvature
is equivalent to a rank 4 bispinor
\begin{equation}
\label{appendixc6}
\left(\begin{array}{c}\zeta^{abcd}\\
\omega_{\dot a\dot b\dot c\dot d}\end{array}\right)
\end{equation}
such that
\begin{eqnarray}
\label{appendixc7}
\zeta^{abcd}=\zeta^{bacd}&=&\zeta^{abdc}=\zeta^{cdab},
\\
\label{appendixc8}
\omega_{\dot a\dot b\dot c\dot d}
=\omega_{\dot b\dot a\dot c\dot d}
&=&\omega_{\dot a\dot b\dot d\dot c}
=\omega_{\dot c\dot d\dot a\dot b}\,,
\\
\label{appendixc9}
\zeta^{ab}{}_{ab}&=&0,
\\
\label{appendixc10}
\omega^{\dot a\dot b}{}_{\dot a\dot b}&=&0.
\end{eqnarray}
Weyl curvature is expressed via the bispinor
(\ref{appendixc6}) as
\begin{multline}
\label{appendixc11}
R^{\kappa\lambda\mu\nu}=
(\gamma^0\gamma^2\gamma^\kappa\gamma^\lambda)_{ab}
\,\zeta^{abcd}
\,(\gamma^0\gamma^2\gamma^\mu\gamma^\nu)_{cd}
\\
+(\gamma^0\gamma^2\gamma^\kappa\gamma^\lambda)^{\dot a\dot b}
\,\omega_{\dot a\dot b\dot c\dot d}
\,(\gamma^0\gamma^2\gamma^\mu\gamma^\nu)^{\dot c\dot d}\,.
\end{multline}
Note that the spinor conditions (\ref{appendixc9}), (\ref{appendixc10})
are needed to ensure the fulfillment of the tensor conditions
(\ref{appendixc4}), (\ref{appendixc5}).
Note also that ${}^*\!R=R^*=iR$ if and only if $\omega=0$,
and ${}^*\!R=R^*=-iR$ if and only if $\zeta=0$.

We say that the complex Weyl curvature $R$
factorizes if
$R_{\kappa\lambda\mu\nu}=F_{\kappa\lambda}F_{\mu\nu}$
for some antisymmetric rank 2 tensor $F$.
We say that the spinor $\zeta$ in the bispinor(\ref{appendixc6})
factorizes if
$\zeta^{abcd}=\xi^a\xi^b\xi^c\xi^d$
for some rank 1 spinor $\xi^a$.
We say that the spinor $\omega$ in the bispinor (\ref{appendixc6})
factorizes if
$\omega_{\dot a\dot b\dot c\dot d}=
\eta_{\dot a}\eta_{\dot b}\eta_{\dot c}\eta_{\dot d}$
for some rank 1 spinor $\eta_{\dot a}$.
Examination of formulae (\ref{appendixc2})--(\ref{appendixc11})
establishes the following

\begin{lemma}
\label{factorizationofcurvature}
A complex Weyl curvature factorizes if and only if
one of the spinors in the bispinor (\ref{appendixc6}) factorizes
and the other is zero.
\end{lemma}

We see that a factorized complex Weyl curvature is equivalent
to a rank 1 spinor $\xi^a$ or $\eta_{\dot a}$.
This spinor is, effectively, the fourth root of curvature,
and is determined uniquely up to multiplication by
$i^n$, $n=0,1,2,3$.

\end{document}